\documentclass{article}
\usepackage{psfig}
\usepackage{rotating}
\usepackage{emulateapj}
\usepackage{apjfonts}

\makeatletter
\let\@internalcite\cite
\def\cite{\def\astroncite##1##2{##1\ ##2}\@internalcite}
\def\citey{\def\astroncite##1##2{##1\ (##2)}\@internalcite}
\def\@citex[#1]#2{\if@filesw\immediate\write\@auxout{\string\citation{#2}}\fi
  \def\@citea{}\@cite{\@for\@citeb:=#2\do
    {\@citea\def\@citea{; }\@ifundefined
       {b@\@citeb}{{\bf ??}\@warning
       {Citation `\@citeb' on page \thepage \space undefined}}%
{\csname b@\@citeb\endcsname}}}{#1}}
\def\@cite#1#2{#1\if@tempswa #2\fi}
\def\@biblabel#1{}
\def\astroncite#1#2{#1\ #2}
\makeatother

\def\rxte{{\em RXTE}}
\def\pca{{\em PCA}}
\def\asm{{\em ASM}}

\def\asca{{\em ASCA}}
\def\euve{{\em EUVE}}
\def\mcg{{MCG$-$6-30-15}}
\def\cyg{{Cyg~X-1}}
\def\ngc{{NGC~5548}}
\def\msun{{\rm M}_\odot}

\def\aproxgt{\mathrel{%
      \rlap{\raise 0.511ex \hbox{$>$}}{\lower 0.511ex \hbox{$\sim$}}}}
\def\aproxlt{\mathrel{%
      \rlap{\raise 0.511ex \hbox{$<$}}{\lower 0.511ex \hbox{$\sim$}}}}

\begin{document}

\slugcomment{Submitted 1999 June 23; Accepted 2000 January 12}

\lefthead{Nowak \& Chiang}
\righthead{Implications of the X-ray Variability for the Mass of \mcg}

\title{Implications of the X-ray Variability for the Mass of \mcg}

\author{Michael A. Nowak\altaffilmark{1} and James Chiang\altaffilmark{1}}

\altaffiltext{1}{JILA, University of
  Colorado, Campus Box 440, Boulder, CO~80309-0440, USA;
  mnowak@rocinante.colorado.edu, chiangj@panza.colorado.edu}   

\received{1999 June 24}
\accepted{1999 September 21}

\begin{abstract}
  The bright Seyfert 1 galaxy \mcg\ shows large variability on a variety of
  time scales.  We study the $\aproxlt 3$\,day time scale variability using
  a set of simultaneous archival observations that were obtained from
  \rxte\ and the {\it Advanced Satellite for Cosmology and Astrophysics}
  (\asca). The \rxte\ observations span nearly $10^6$\,sec and indicate
  that the X-ray Fourier Power Spectral Density has an rms variability of
  16\%, is flat from approximately $10^{-6}$--$10^{-5}$\,Hz, and then
  steepens into a power law $\propto f^{-\alpha}$ with $\alpha\aproxgt 1$.
  A further steepening to $\alpha \approx 2$ occurs between
  $10^{-4}$--$10^{-3}$\,Hz.  The shape and rms amplitude are comparable to
  what has been observed in \ngc\ and \cyg, albeit with break frequencies
  that differ by a factor of $10^{-2}$ and $10^{4}$, respectively. If the
  break frequencies are indicative of the central black hole mass, then
  this mass may be as low as $10^6~{\rm M}_\odot$.  An upper
  limit of $\sim 2$\,ks for the relative lag between the 0.5--2\,keV \asca\ 
  band compared to the 8--15\,keV \rxte\ band was also found. Again by
  analogy with \ngc\ and \cyg, this limit is consistent with a relatively
  low central black hole mass.
\end{abstract}




\section{Introduction}\label{sec:intro}

The type 1 Seyfert galaxy \mcg\ has in recent years been the subject of
intense study owing to the discovery by the {\em Advanced Satellite for
  Cosmology and Astrophysics} (\asca) of a resolved, broad iron K$\alpha$
fluorescent line in its hard X-ray spectrum (\cite{tanaka:95b}).  The shape
of the line is consistent with a gravitationally and Doppler shifted
emission line which originates from near the inner edge of an accretion
disk around a black hole.  In the X-rays, \mcg\ is also one of the brighter
and more variable type 1 Seyferts.  It is therefore hoped that by examining
the correlated variability between the ionizing X-ray continuum and various
components of the Fe K$\alpha$ line, one can obtain a size scale for the
system and thence a mass for the black hole (\cite{reynolds:99a}).  This
requires, however, that the various components of the line be spectrally
resolved on time scales shorter than the intrinsic response time scale of
the line-emitting material.

At present, \asca, {\em BeppoSAX}, and {\em Chandra} are the only
X-ray telescopes with the spectral resolution to measure fluxes
in different line components separately.  Unfortunately, the integration
times ($\ga 10$\,ks) required to obtain this resolution are longer than the
light travel time across the inner edge of an accretion disk around a
$10^8\,M_\odot$ Schwarzschild black hole.  Furthermore, the fits to the
iron line appear to require a Kerr geometry (although see
\cite{reynolds:97b}) in order to explain the broad red wing of the line and
the lack of significant emission blueward of 6.4\,keV, thus making the
relevant time scales even shorter.  Nonetheless, over longer time scales,
\asca\ has measured significant variability in the shape of the \mcg\ 
Fe~K{$\alpha$} line (\cite{iwasawa:96a}).

Time-resolved spectral investigations (\cite{iwasawa:96a,iwasawa:99a})
suggest that the broad and narrow components of the iron line are
correlated differently with the flux state depending on the time scale
investigated.  For integrations $\ga 10^4$\,s, the narrow component varies
with the continuum flux whilst the broad component appears to be
anti-correlated.  In contrast, on shorter time scales the broad component
responds immediately to flux changes whilst the narrow component remains
constant.  \citey{iwasawa:99a} suggest that multiple, localized X-ray
flares occur on the disk surface near the inner edge and illuminate only a
relatively small region of the disk.  These small regions make
contributions to narrow ranges in line redshift and thus produce very
complex temporal behavior.

\begin{figure*}
\centerline{
\psfig{figure=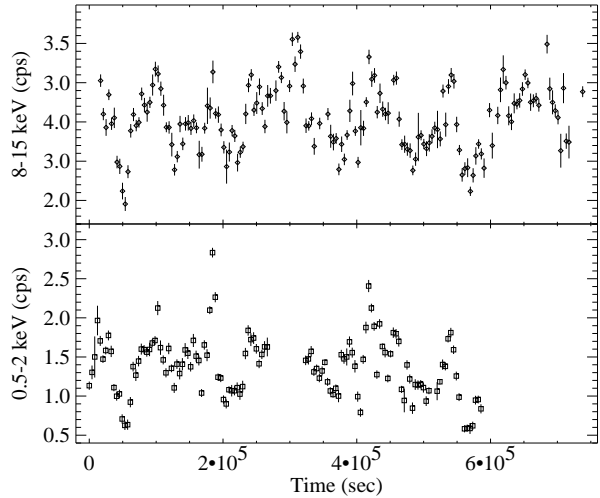,width=0.45\textwidth}
}
\caption{\small 
  The \rxte\ 8--15\, keV band (top) and the \asca\ 0.5--2\,keV band
  (bottom) lightcurves, in 4096\,s timebins. These data have been obtained
  from the archival 1997 August simultaneous \asca/\rxte\ observation
  (Lee et al. 1998).  
  \protect{\label{fig:curves}}}
\end{figure*}

In support of this interpretation, \citey{iwasawa:99a} consider the iron
line associated with a very short, bright flare from \mcg\ observed during
1997 August by \asca\ (see Fig.~\ref{fig:curves}).  The line was very
redshifted and had little or no emission blueward of 6\,keV.  Its shape was
consistent with having originated entirely from a small region at $r \simeq
5~ {\rm GM/c^2}$ on the approaching side of the disk.  The flare lasted
about 4\,ks, which is also approximately the orbital period at this radius
for a $10^7\,\msun$ black hole.  In order for the line not to be
significantly more smeared by the orbital motion, \citey{iwasawa:99a}
estimate a black hole mass of $\sim 2 \times 10^8\,\msun$. Alternatively,
they suggest that a much lower mass is possible if all the line emission arises
from $r < 5~{\rm GM/c^2}$.

A major complication that a high-mass model faces is the large amplitude
and rapid X-ray variability of \mcg\ (Fig.~\ref{fig:curves}).  In
particular, during the performance verification phase of \asca,
\citey{reynolds:95a} noted that the 0.5--10\,keV flux increased by a factor
1.5 over 100\,s, i.e., the dynamical time scale at the inner disk edge
surrounding a maximal Kerr, $2\times10^6\,\msun$ black hole.  Typically one
expects the bulk of the X-ray variability to occur on dynamical time scales
or longer.  It is clearly important to determine whether this variability
represented a rare, rapid event or if such time scales are truly
characteristic of the behavior of \mcg.

One might expect that characteristic time scales should scale with the mass
of the central object (see, e.g., the discussions of \cite{mchardy:88a} and
\cite{edelson:99a}); therefore, in this work we try to gauge the size of
the system and the mass of the black hole in \mcg\ by studying its
characteristic X-ray variability properties in comparison to other black
hole systems such as \cyg\ and \ngc. For the high-frequency variability
analysis, we use the 1997 August simultaneous \asca/{\em Rossi X-ray Timing
  Explorer} (\rxte) observation discussed by \citey{iwasawa:99a} (see also
\cite{lee:98a}).  In our analysis we screen the \asca\ data
as outlined by \citey{brandt:96a}, except that we use the more
stringent criteria of 7~GeV/c for the rigidity and an elevation angle of
$10^\circ$.  Data from both SIS detectors are combined into a single
lightcurve.  For the \rxte\ data, we use screening criteria and analysis
techniques appropriate for faint sources, as we have previously discussed
in \citey{chiang:99a}.  Specifically, we only analyze top layer data from
proportional counter units 0, 1, and 2.

\section{Power Spectra}\label{sec:psd}

\mcg\ is detected by the {\em All Sky Monitor} (\asm; \cite{levine:96a}) on
\rxte\ with a mean count rate of 0.44\,cps in the 1.3--12.2\,keV band.
Currently, there is a 0.1 cps uncertainty in the zero level offset, as well
as comparable magnitude systematic time-dependent variations due to, for
example, the solar angular position relative to the source (Remillard 1999,
priv. comm.).  Assessing the very low frequency ($f \aproxlt 10^{-6}$\,Hz)
X-ray variability of \mcg\ is therefore problematic.  Efforts are currently
underway, however, to revise the \asm\ data reduction process in order to
minimize such systematic effects (Remillard 1999, priv. comm.); therefore,
an ultra-low frequency variability study of \mcg\ in principle will be
feasible in the near future.

\begin{figure*}
\centerline{
\psfig{figure=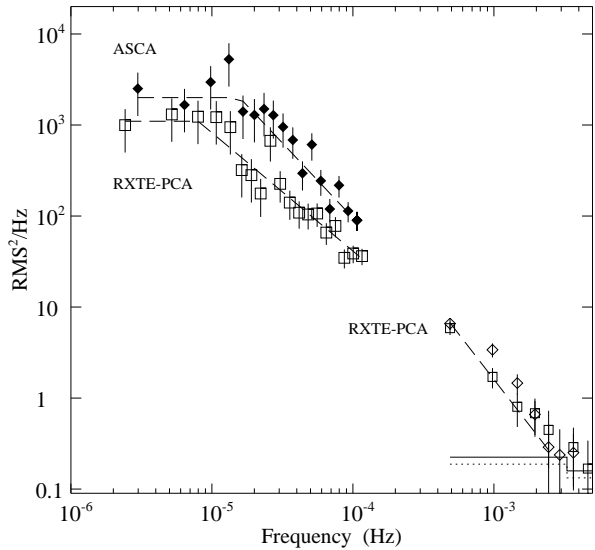,width=0.45\textwidth}
}
\caption{\small 
PSD constructed from 
  \asca\ 0.5--2\,keV (filled diamonds), \rxte\ 1.8--3.6\,keV (open
  diamonds), and \rxte\ 8--15\,keV (open squares) lightcurves.  Dashed
  lines show the broken power law fits discussed in the text and an
  $f^{-2}$ power law. Background and noise have been subtracted from the
  high-frequency 1.8--3.6\,keV and 8--15\,keV \rxte\ PSD, with the residual
  noise levels shown as a dotted and solid line, respectively.
  \protect{\label{fig:psd}}}
\end{figure*}

We are able, however, to investigate the high-frequency power spectral
density (PSD) of \mcg\ by using the \rxte\ 8--15\,keV and the \asca\ 
0.5--2\,keV lightcurves binned on 4096\,sec time scales.  Both lightcurves
contain data gaps, especially on the orbital time scale of $\approx 5$\,ks
due to blockage by the Earth and passage through the South Atlantic
Anomaly. We therefore use the techniques of \citey{lomb:76a} and
\citey{scargle:82a} to calculate the PSD, and we only consider frequencies
$f \aproxlt 10^{-4}$\,Hz (evenly sampled in intervals of the inverse of the
observation duration).  Higher time resolution lightcurves showed excess
power on the $\approx 5$\,ks orbital time scale.  The results, binned over
the greater of four contiguous frequency bins or logarithmically over
$f\rightarrow 1.15~f$, are presented in Fig.~\ref{fig:psd}.  Here we use a
one-sided normalization where integrating over positive frequencies yields
the total mean square variability relative to the squared mean for the
particular lightcurve being analyzed. The \asca\ lightcurve shows 28\% rms
variability, whereas the \rxte\ lightcurve shows 16\% rms variability.
Both PSD have comparable shapes, i.e., flat from $\approx
10^{-6}$--$10^{-5}$\,Hz and slightly steeper than $f^{-1}$ at
higher frequencies.

We have fit a broken power law to the data, assuming error bars equal to
the average PSD value divided by the square root of the number of frequency
bins averaged over.  Although such error estimates are only valid for
averages made from independent frequency bins (which is not strictly true
for the Lomb-Scargle periodogram; \cite{scargle:82a}), this should provide
a rough estimate, especially as the lightcurves are nearly evenly sampled.
The best fit PSD slopes are  $-1.3\pm0.2$ for the \pca\ and
$-1.6\pm0.3$ for \asca\ (errors are $\Delta\chi^2=2.7$).  The break
frequencies are found to be $(8\pm3)\times10^{-6}$\,Hz for the \pca,
and $(1.5\pm0.5)\times10^{-5}$\,Hz for \asca.  These break
frequencies are consistent with the value found by \citey{mchardy:98a}.

At frequencies $> 5\times10^{-4}$\,Hz there are enough contiguous data
segments to be able to calculate the PSDs using standard FFT methods
(\cite{nowak:99a}, and extensive references therein). In Fig.~\ref{fig:psd}
we show the calculated, background and noise-subtracted, PSD for the
1.8--3.6\,keV and 8--15\,keV \rxte\ bands.  Below $\approx
2\times10^{-3}$\,Hz, where signal to noise is greatest, both PSDs are
consistent with being $\propto f^{-2\pm0.3}$ and having 6\% rms variability.
From the generated background lightcurves, we estimate that background
fluctuations contribute at most 1.5\% and 2\% rms variability,
respectively, to these PSDs at $<2\times10^{-3}$\,Hz.  As the PSDs of the
background lightcurves tend to be slightly steeper than $f^{-2}$, there may
be some trend for background fluctuations to steepen the observed
high-frequency PSDs, and, in fact, \citey{yaqoob:97a} find a slightly
flatter ($\propto f^{-1.4}$) high-frequency PSD for \mcg.

\section{Time Delays}\label{sec:delay}

In order to examine time delays between the low energy \asca\ lightcurve
and the high energy \rxte\ lightcurve, we use the Z-transformed Discrete
Cross-Correlation Function (ZDCF) of \citey{alexander:97a}, which is based
upon the DCF method of \citey{edelson:88a}.  Auto-correlation functions can
also be computed by this procedure, and we note that PSD derived from
autocorrelations calculated via the ZDCF yield identical results to those
presented in Fig.~\ref{fig:psd}.  We use the Monte Carlo methods described
by \citey{peterson:98a} to assess the significance of any uncertainties due
to flux measurement errors as well as uneven or incomplete sampling.

\begin{figure*}
\centerline{
\psfig{figure=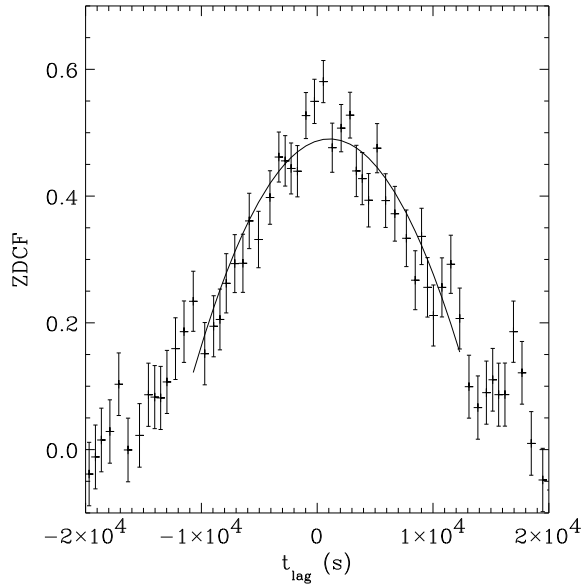,width=0.45\textwidth}
}
\caption{\small Z-transformed discrete correlation function (ZDCF) of the
  SIS, 0.5-2\,keV lightcurve relative to the 8--15\,keV \rxte-\pca\
  lightcurve (512\,s time bins).  The solid line is a parabola fit to
  the ZDCF values.  Monte Carlo simulations of these data yield an
  upper-limit on the \rxte\ time lag of $2.2$\,ks (90\% Confidence
  Level).  \protect{\label{fig:cross}}}
\end{figure*}

We have previously applied these methods to a simultaneous {\em Extreme
  Ultraviolet Explorer} (\euve)/\asca/\rxte\ observation of \ngc\ 
(\cite{chiang:99a}), where we found evidence for the low energy \asca\ band
leading the high energy \rxte\ band by 5\,ks.  The results for the \mcg\ 
data are shown in Fig.~\ref{fig:cross}, for which we have used lightcurves
binned at 512\,s resolution.  A positive delay indicates that the \rxte\ 
light curve lags the \asca\ light curve.  Due to the ambiguities associated
with interpreting cross-correlation results, particularly for such small
delays, we considered three different measures of the ``lag''.  For each
Monte Carlo trial, we estimate the characteristic lag by (1) fitting a
parabola to the ZDCF values to find the location of the peak, (2) computing
the centroid of the ZDCF over positive values bracketing the maximum value,
and (3) using the location of the actual maximum value of the ZDCF.  In all
three cases, our simulations yield evidence for a positive lag at various
degrees of significance: $\tau_{\rm fit} = 0.9 \pm 0.7\,$ks, $\tau_{\rm
  centroid} = 0.9 \pm 1.3\,$ks, $\tau_{\rm max} = 0.4^{+1.9}_{-1.2}$ (90\%
C.L.).  Although the fitted peak estimate is consistent with a positive
lag, both the centroid and ZDCF-maximum estimates are formally consistent
with zero giving an upper limit of $\tau \aproxlt 2\,$ks.

At high Fourier frequency, we have used the 1.8--3.6\,keV and 8--15\,keV
\rxte\ lightcurves discussed above to search for frequency-dependent time
lags using standard FFT techniques. (A complete discussion of such methods,
including calculation of error bars, is presented in \cite{nowak:99a} and
references therein).  No significant time delays were found, and the
1-$\sigma$ upper limits were 50--100\,s in the
$5\times10^{-4}$--$2\times10^{-3}$\,Hz range.  Note that there were an
insufficient number of uninterrupted, strictly simultaneous \asca/\rxte\ 
lightcurves to allow calculation of their relative time delays via direct
FFT methods.

\section{Discussion}\label{sec:discuss}

The \mcg\ PSD breaks to being $\propto f^{-1}$ at $\approx 10^{-5}$\,Hz,
and then breaks to being approximately $\propto f^{-2}$ between
$10^{-4}$--$10^{-3}$\,Hz.  This is to be compared to the \cyg\ PSD which
has a comparable rms amplitude and shape, and has a set of PSD breaks at
frequencies between 0.03--0.3\,Hz and 1--10\,Hz (\cite{nowak:99a}, and
references therein). The black hole mass in \cyg\ is estimated to be
$10\,\msun$ (\cite{herrero:95a}); therefore, if these break frequencies
scale with mass, then the central black hole mass of \mcg\ could be as low
as $10^6\,\msun$.  \ngc, which is believed to have a central black hole
mass of $10^8\,\msun$ (\cite{done:96a,chiang:96a,peterson:99a}), shows a
similar PSD with break frequencies at $\approx 6 \times 10^{-8}$\,Hz and
between $3\times10^{-7}$--$3\times10^{-6}$\,Hz (\cite{chiang:99a}).  Again,
if the break frequencies scale with mass, then the central black hole mass
for \mcg\ could be several orders of magnitude lower than that for \ngc.

To date, the most carefully studied X-ray PSD for any AGN is that for
NGC~3516 (\cite{edelson:99a}). (See also \cite{mchardy:98a} who present
preliminary results for a number of AGN, including \mcg.) For NGC~3516 the
PSD was seen to break from nearly flat at $\aproxlt 3\times 10^{-7}$\,Hz,
and then gradually steepen into an $f^{-1.74}$ power law up to frequencies
as high as $\approx 10^{-3}$\,Hz.  NGC~3516 is seen to be intermediate
between \ngc\ and \mcg.  Based upon these measurements (and upon other
factors, such as the source luminosity), \citey{edelson:99a} argue for a
black hole mass in the range of $10^7$--$10^8\,\msun$, i.e., intermediate
to the masses of NGC~5548 and \mcg\ discussed above.

The \rxte/\asca\ lag upper limit for \mcg\ is also intermediate between the
observed X-ray lags for \cyg\ (\cite{miyamoto:88a,nowak:99a}) and NGC~5548
(\cite{chiang:99a}). The time lag observed in NGC~5548, effectively
measured on the $f^{-2}$ portion of its PSD, is 5\,ks. Time lags on the
flat portion of the \ngc\ PSD could be considerably longer
(\cite{chiang:99a}). Near the PSD break from flat to $f^{-1}$, the X-ray
time lags in \cyg\ are $\approx 0.1$\,s, whilst on the $f^{-2}$ portion
of the PSD the X-ray time lags are $10^{-3}$--$10^{-2}$\,s.  The \mcg\ time
lags may cover a similar dynamic range from $<2$\,ks (overall lag) to
$<100$\,s (high frequency lag\footnote{Although the low energy band here
  differed from the \asca\ band, if the time lags scale logarithmically
  with energy (\cite{nowak:99a} and references therein) we would have
  expected the lag with respect to the \asca\ band to be approximately a
  factor of 2 greater.}).

If the characteristic variability and lag times are indicative of mass,
then a mass as low as $10^6\,\msun$ may be required for the central black
hole of \mcg.  Assuming a bolometric luminosity of $4\times 10^{43}~{\rm
  ergs~s^{-1}}$ (\cite{reynolds:97a}), this would imply that \mcg\ is
emitting at $30\%$ of its Eddington rate, which is large but still
plausible.  A relatively low central black hole mass would make the large
amplitude, rapid variability reported by \citey{reynolds:95a} much easier
to understand, whereas a mass as large as the $2 \times 10^8\,\msun$, the
upper end discussed by \citey{iwasawa:99a}, seems very unlikely.  The
$\aproxlt 100$\,s lags seen at high frequency then likely provide an upper
limit to the Compton diffusion time scale (see the discussion of
\cite{nowak:99a}). These time scales are problematic for future hopes of
simultaneously temporally and spectrally resolving the iron K$\alpha$ line
in this system with {\em Constellation-X} as it will require $\approx
1$\,ks integration times to study the line profile of \mcg\ 
(\cite{young:99a}).

A 30\% Eddington luminosity indicates that it is worthwhile to search for
ultra-low frequency ($f \aproxlt 10^{-6}$\,Hz) variability in excess of
$\approx 3\%$ rms, that is, above an extrapolation of the flat part of the
\asca\ and \rxte-\pca\ PSDs.  `Very high state', i.e., $\aproxgt {\cal
  O}(30\%~L_{\rm Edd})$, galactic black hole candidates can show PSDs that
are flat below $\approx 10^{-2}$\,Hz, are approximately proportional to
$f^{-2}$ PSD between $\approx 10^{-2}$--$10^{-1}$\,Hz, are flat again
between $\approx 10^{-1}$--$1$\,Hz, and then break into an $f^{-2}$ PSD at
higher frequencies. (Specifically, see Fig.~4b of \cite{miyamoto:91a},
which shows a `very high state' PSD of GX339$-$4.)  The low frequency
portion of the `very high state' PSD has no simple analogy in the (usually
observed) low/hard state of \cyg, where ultra-low frequency noise is
typically associated with dipping activity due to obscuration by the
accretion stream (\cite{angelini:94a}).  Previous models, for example, have
associated `very high state' low-frequency variability with fluctuations of
a viscously unstable $\alpha$-disk (\cite{nowak:94a}, and references
therein).

This highlights the major caveat that needs to be mentioned; we do not know
the average PSD shape nor the scaling of the break frequencies as a
function of fractional Eddington luminosity in either galactic black hole
candidates or AGN.  This is an especially important consideration as the
mass estimates discussed above imply a large range of Eddington luminosity
ratios.  Considering all the evidence for rapid variability and extremely
short time lags in \mcg\ discussed above, however, a low mass for \mcg\ 
seems to us very compelling.  With the advent of the {\em X-ray Multiple
  Mirror} ({\em XMM}) mission, which has large effective area and is
capable of extremely long, uninterrupted observations, these analyses will
become more detailed for \ngc\ and \mcg, and will allow one to develop a
statistical sample of numerous other AGN.

\acknowledgements We thank R. Remillard for generating an \asm\ lightcurve
of \mcg, and O. Blaes, K. Pottschmidt, N. Murray, and J. Wilms for useful
conversations.  This work has been financed by NASA Grants NAG5-4731,
NAG5-7723, and NAG5-6337.  This research has made use of data obtained
through the High Energy Astrophysics Science Archive Research Center Online
Service, provided by the NASA/Goddard Space Flight Center.

\end{document}